\documentclass[letter]{aa} 

\usepackage{graphicx}
\usepackage{comment}
\usepackage{hyperref}
\hypersetup{colorlinks=true,linkcolor=[rgb]{1.,0.2,0.2},citecolor=[rgb]{0.1,0.4,1.},filecolor=[rgb]{0.7,0.2,0.2},urlcolor=[rgb]{0.7,0.2,0.2}}

\usepackage{txfonts}
\begin{document}

   \title{KMHK~1762: Another star cluster in the Large Magellanic Cloud age gap}

   \subtitle{}

   \author{M. Gatto
         \inst{1} \thanks{massimiliano.gatto@inaf.it}
         \and
         V. Ripepi\inst{1}
         \and 
         M. Bellazzini\inst{3}
         \and
         M. Tosi\inst{3}
         \and
         C. Tortora\inst{1}
         \and
         M. Cignoni\inst{1,4,5}
         \and
         M. Dall'Ora\inst{1}
         \and
         M.-R. L. Cioni\inst{6}
         \and
         F. Cusano\inst{3}
         \and
         G. Longo\inst{2}
         \and
         M. Marconi\inst{1}
         \and
         I. Musella\inst{1}
         \and
         P. Schipani\inst{1}
         \and
         M. Spavone\inst{1}
         }

   \institute{INAF--Osservatorio Astronomico di Capodimonte, Via Moiariello
 16, 80131, Naples, Italy 
 \and
 Dept. of Physics, University of Naples Federico II, C.U. Monte Sant'Angelo, Via Cinthia, 80126, Naples, Italy
 \and
 INAF--Osservatorio di Astrofisica e Scienza dello Spazio, Via Gobetti 93/3, I-40129 Bologna, Italy
 \and
 Physics Departement, University of Pisa, Largo Bruno Pontecorvo, 3, I-56127 Pisa, Italy
 \and
 INFN, Largo B. Pontecorvo 3, 56127, Pisa, Italy
 \and
 Leibniz-Institut f\"ur Astrophysik Potsdam, An der Sternwarte 16, D-14482 Potsdam, Germany
             }

   \date{
   }

 
  \abstract
   {The star cluster (SC) age distribution of the Large Magellanic Cloud (LMC) exhibits a gap from $\sim$ 4 to 10 Gyr ago, with an almost total absence of SCs. Within this age gap, only two confirmed SCs have been identified hitherto. Nonetheless, the star field counterpart does not show the same characteristics, making the LMC a peculiar galaxy where star formation history and cluster formation history appear to differ significantly.}
   {We re-analyzed the color-magnitude diagram (CMD) of the KMHK 1762 SC by using  the deep optical photometry  provided by the ``Yes, Magellanic Clouds Again'' survey, to robustly assess its age.}
   {First, we partially removed foreground and/or field stars by means of parallaxes and proper motions obtained from the {\it Gaia} Early Data Release 3.
   Then, we applied the Automated Stellar Cluster Analysis package to the cleaned photometric catalogue to identify the isochrone that best matches the CMD of KMHK 1762.}
   {The estimated age of KMHK 1762 is $\log (t) = 9.74 \pm 0.15$ dex ($\sim$5.5 Gyr), that is more than 2 Gyr older than the previous estimation which was obtained with shallower photometry. This value makes KMHK 1762 the third confirmed age gap SC of the LMC.}
   {The physical existence of a quiescent period of the LMC SC formation is questioned. We suggest it can be the result of an observational bias, originated by the combination of shallow photometry and limited investigation of the LMC periphery.}


   \maketitle
%

\section{Introduction}

The Large Magellanic Cloud (LMC) is the most massive satellite of the Milky Way (MW) and because of its proximity \citep[i.e. 50~kpc,][]{deGrijs&Wicker&Bono2014} it is possible to obtain color-magnitude diagrams (CMDs) deeper than the main-sequence turn-off (MSTO) of the oldest stellar population present in the galaxy.  
The LMC is therefore one of the few galaxies where it is feasible to carry out a detailed analysis of its star cluster (SC) system \citep[e.g.,][]{Pietrzynski&Udalski2000,Glatt-2010,Piatti-2014,Pieres-2016,Nayak-2016} or to derive its star formation history (SFH) in great detail by studying its resolved stellar population \citep[e.g.][]{Harris&Zaritsky2004,Harris&Zaritsky2009,Cignoni-2013,Weisz-2013,Rubele-2012, Mazzi-2021, Piatti&Geisler2013}.
Investigating if the SFHs of the SCs and field population agree to each other would have important consequences in our understanding of galaxy formation and evolution. 
For example, in galaxies well beyond the Local Group, where it is impossible to resolve individually the field stars, their SC system might be adopted as a proxy to study the galaxy's SFH.\par
The LMC presents a well known ``age-gap'' in the SC age distribution, in the range from about 4 to 10 Gyr \citep{Jensen-1988,DaCosta1991} where until a few years ago only 1 member was known, namely the SC ESO121-03 with an estimated age of $9.0 \pm 0.8$~Gyrs \citep[][but see also \citealt{Mateo-1986}]{Mackey-2006}.
On the other hand, several works from different authors provided evidence that the stellar field counterpart shows a significant population formed in that age range \citep[e.g.][]{Tosi2004,Piatti&Geisler2013,Mazzi-2021}, making the LMC a peculiar galaxy where the SF activity of SCs and field are different during the age gap period, whereas they are strongly similar outside this interval.
As it is thought that most stars of a galaxy form in SCs \citep{Lada&Lada2003}, if the age gap is a real feature of the LMC, the question we should answer now is: ``What did happen to all SCs that formed the observed field stars in the LMC during the age gap?''.\par
In the most recent years, and particularly during the last decade, the number of projects surveying different regions of the LMC to search for new SCs has been remarkably increased, leading to the discovery of hundreds of new SCs in the MCs \citep{Sitek-2016,Piatti-2015a,Piatti-2018,Piatti2021a}.
Most of these works targeted only the central regions of the LMC, where confusion due to crowding makes it extremely difficult to obtain deep and accurate photometry, needed to unambiguously detect old SCs, thus bringing modest advances in the comprehension of the age gap feature.
One of the few exceptions was the extensive research made by \citet{Pieres-2016}, that probed the outer Northern side of the LMC through the public Dark Energy Survey \citep[DES,][]{DES-Abbott-2016} data, up to distances of 10 kpc from the LMC centre, which increased the number of known SCs in those observed fields by more than 40\%.
They also provided an estimate of their main parameters, i.e. age, metallicity, distance modulus and reddening, for a sub-sample of 117 SCs by adopting a maximum-likelihood approach to estimate the relevant SC parameters. Besides the confirmation of ESO121-03 as a genuine member of the age gap, \citet{Pieres-2016} also revisited the age of NGC~1997 to be $\sim 4.5 \pm 0.1$~Gyrs, about 2 Gyrs older than previous evaluations \citep{Piatti-2009,Palma-2016}, making it potentially a further age gap member, as the youngest gap edge is not well defined.\par 
Recently, \citet[][G20 hereafter]{Gatto-2020} presented the discovery of 78 new candidate SCs in the outskirts of the LMC, 16 of which with estimated ages falling within the gap. These SCs were detected thanks to two VST surveys: ``SMC in Time: Evolution of a Prototype interacting late-type dwarf galaxy'' \citep[STEP,][]{Ripepi-2014} and the first 21 sq. deg analysed of the ``Yes, Magellanic Clouds Again (YMCA - Gatto et al., in prep.) survey. 
\citet[][]{Piatti2021b} re-analyzed a sub-sample of age gap SC candidates reported in G20, using data from the SMASH survey \citep[][]{Nidever-2017}. 
They concluded that some of the candidate SCs may not be real physical systems, but rather field star density fluctuations. A definitive assertion of the real nature of those SC candidates can be obtained only with deeper photometric observations.
Very recently, \citet{Piatti2022b} re-estimated the age of the LMC SC KMHK~1592 to be $8.0 \pm 0.5$~Gyrs through deep photometric observations carried out with the GEMINI South. This is the second LMC genuine SC well within the age gap. Because of the presence of only two confirmed members within the age gap\footnote{
\citet{Piatti2022b} did not mention NGC\,1997 as an age gap member.}
\citet{Piatti2022b} favored the scenario of a capture from an external galaxy, like for example the SMC. As \citet{Piatti2022b} suggested, if other confirmed age gap SCs were discovered, the in-situ origin hypothesis would be re-inforced. Hence, it is crucial to detect as many as possible age gap members, if they exist.\par
In this context, we are carrying out a systematic search of unknown SCs and a detailed re-analysis of the already catalogued SCs in the YMCA tiles not investigated in G20.
During this work, we came across the already known SC KMHK~1762 \citep[also referred to as OHSC~37 in the catalog by][]{Bica-2008}, whose CMD attracted our attention, as it showed at first sight a potentially older age than what is known from the literature. 
While the candidate SCs detected in G20 should still be confirmed with deeper photometry, the SC nature of KMHK~1762 is secured by some favourable characteristics: i) it clearly stands out above the background as it resides in a very low-density environment; ii) it is relatively populous; iii) it has several evolved stars which can be confirmed members of the SC and that make it easier to identify the different evolutionary phases.
Previous photometric observations of KMHK~1762 were carried out at the Cerro Tololo Inter-American Observatory (CTIO) with the 0.9~m telescope \citep[][]{Geisler-1997}. 
These authors adopted the magnitude difference in the Washington $T_1$ filter between the red clump (RC) stars and the MSTO ($\delta T_1$) to derive the age of KMHK~1762, obtaining $t \sim 2.7$~Gyrs. A metal content of $[{\rm Fe/H}] = -0.91$, based on the spectroscopic measurements of the Ca{\tt II} triplet for one spectroscopically confirmed member star \citep[][]{Olszewski-1991}, placed KMKH~1762 at a fairly lower metallicity level with respect to other SCs with similar ages \citep[][]{Geisler-1997}.\par
In this letter, we report the result of our new study of KMHK~1762 SC, based on the YMCA deep and accurate photometry, complemented with parallaxes and proper motions (PMs) from the {\it Gaia} Early Data Release 3 \citep[EDR3;][]{Gaia-Brown-2021}.

\section{Observations and data reduction}
\label{sec:data}

YMCA is an optical survey carried out with the VLT Survey Telescope \citep[VST,][]{Capaccioli&Schipani2011} as part of the Guaranteed Time Observations (GTO) assigned by the European Southern Observatory (ESO) to the Istituto Nazionale di Astrofisica (INAF).
The VST mounts the OmegaCAM which is a mosaic camera of 32-CCD, $16k\times16k$ detectors and has a field-of-view of 1 deg$^2$ with a pixel scale of 0.214 arcsec/pixel.
KMHK~1762 resides within the YMCA tile 5\_38 centred at ($\alpha,\,\delta$) = (07:12:09.384,-69:34:30.360) J2000, (see G20 for a footprint of the YMCA survey), which was observed in $g$ and $i$ bands, with seeing of 1.46\arcsec and 1.13\arcsec, respectively, during February--March 2020.
We used the Astro-WISE pipeline \citep{McFarland-2013} to execute the pre-reduction, astrometry and stacking of the different exposures in order to obtain a single mosaic image for each filter. 
To obtain the Point Spread Function (PSF) photometry and to calibrate the catalogue, we followed the same procedure as in G20 (see their Sect. 2 for full details).
In particular, we used DAOPHOT IV/ALLSTAR packages \citep[][]{Stetson1987,Stetson1992} to carry out  the PSF photometry, and adopted the local standard stars provided by the AAVSO Photometric All-Sky Survey (APASS) to obtain the absolute photometry of the stars in the tile.
Finally, to filter out extended or spurious sources, we required that the {\it SHARPNESS} parameter  calculated by the DAOPHOT IV package lies in the range $-1.0 \leq {\it SHARPNESS} \leq 0.7$.\par
We also carried out an analysis of the photometric completeness, in particular within the cluster region, to evaluate the impact of crowding effects and poor seeing on the SC physical parameters estimation. 
We performed artificial star tests in both g and i bands to retrieve their photometric completeness, following the same procedure reported in \citet{Ripepi-2014}. The results show that our photometry in the innermost, crowded SC regions is 80\% complete in both bands down to g $\sim$ 21.5 mag and 50\% complete down to g $\sim$ 22.5 mag. In all the other regions it is obviously much more complete.

\section{Analysis}

\begin{figure*}
    \centering
    \includegraphics[width=0.38\textwidth]{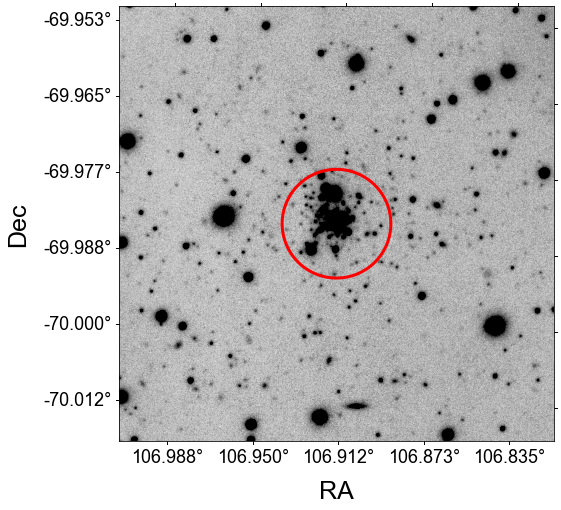}
    \includegraphics[width=0.35\textwidth]{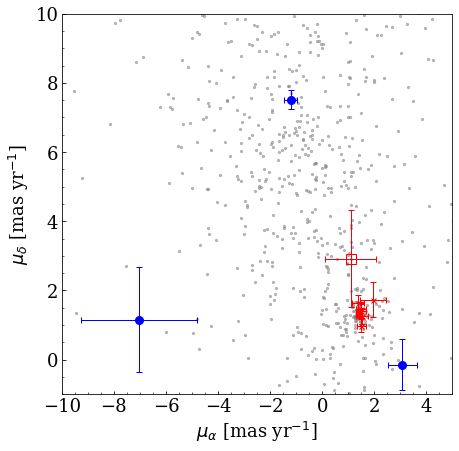}
    \caption{\emph{Left:} Image of about 4\arcmin~in diameter in the $g$-band centred on KMHK~1762. The red circle indicates the SC radius of 0.5\arcmin.
    \emph{Right:} PMs of the 12 stars of KMHK~1762 having astrometric data from {\it Gaia} EDR3. The red crosses are the stars with PMs compatible with the cluster within their uncertainties. The empty red square indicates a star likely with statistically compatible PMs but having large uncertainties (see text). Blue circles are stars with measured PMs beyond $5\sigma$ from the weighted mean SC PM. Grey points represent all stars within 10\arcmin~from the cluster center, without their uncertainties to preserve the readability of the figure.}
    \label{fig:kmhk1762_image}
\end{figure*}

The image of KMHK~1762 is displayed in Fig.~\ref{fig:kmhk1762_image} (left panel). The cluster is located at about 9.8\degr~towards the East of the LMC, making it one of the farthest SCs of the galaxy. 
The centre and the radius of the cluster were determined through the technique developed and described in detail in G20 (see Sect. 3.2), obtaining  $(\alpha, \delta) \simeq (106.9143\degr, -69.984\degr)$ (J2000) and r$\sim$0.5\arcmin.
We expect that LMC and MW field stars are located in front or in the back of KMHK~1762, thus to exploit its CMD with isochrone fitting, we need to first apply a cleaning procedure to mitigate the impact of the contaminant stars.
This task appears to be not straightforward in the case of KMKH~1762, as it is placed in a relatively poorly populated region of the LMC and it is thus difficult to apply the usual procedure to remove contaminants by using the CMD of representative fields around the SC \citep[see for example the procedure by][already adopted in G20]{Piatti&Bica2012}. 
Therefore, to mitigate the stellar field contamination, we took advantage of the recent {\it Gaia} EDR3 data, with the purpose of removing likely MW foreground stars based on their parallaxes and PMs. 
To this aim. we first performed a cross-match of the positions of all YMCA stars within 0.5\arcmin~from the KMHK~1762 centre (i.e. the estimated cluster radius) with the {\it Gaia} EDR3 catalogue, by adopting a maximum tolerance of 1\arcsec, and we obtained 15 stars in common\footnote{The small number of matches is due to the shallow limiting magnitude of {\it Gaia} which is $g \simeq$ 21 mag.}.
Following the criterion described in \citet[][]{Gaia-Luri-2021} we looked for stars not compatible with the LMC distance excluding them if $\varpi > 5~\sigma_{\varpi}$, where $\varpi$ and $\sigma_{\varpi}$ are the {\it Gaia} EDR3 parallaxes and parallax uncertainties. 
Three stars satisfy the previous condition, thus they are likely Galactic foreground stars. The PMs of the remaining 12 common stars are displayed in the right panel of Fig.~\ref{fig:kmhk1762_image}.
We clipped out stars with PMs beyond $5\sigma$ from the weighted mean as they do not likely represent actual SC members. The remaining 9 stars which, based on their kinematics, might belong to the SC possess a weighted mean PM of $(\mu_{\alpha}, \mu_{\delta}) = (1.42 \pm 0.04), (1.31 \pm 0.04)$~mas~yr$^{-1}$.
Among them, the star with $(\mu_{\alpha}, \mu_{\delta}) = (1.09, 2.92)$ ~mas~yr$^{-1}$ is barely consistent with the clump of objects closely piling up around the average KMHK~1762 PM values, and only in virtue of the large uncertainties of its PMs. We mark it with a different symbol in Fig.~\ref{fig:kmhk1762_image}.\par
\begin{figure}
    \hspace*{-0.5cm}
    \includegraphics[width=0.5\textwidth]{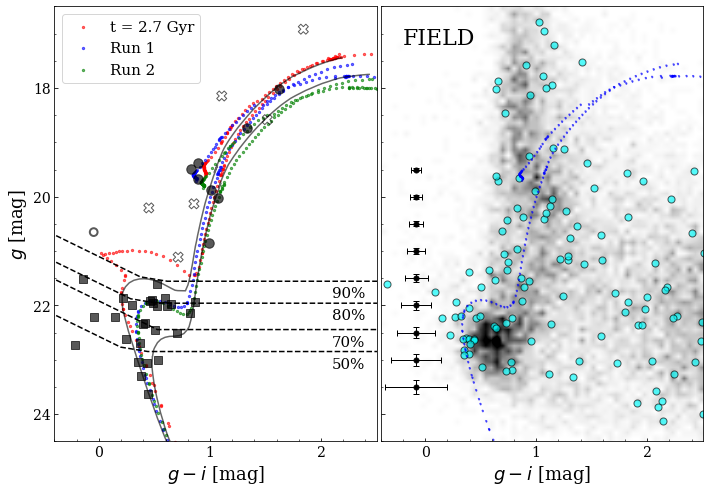}
    \caption{\emph{Left:} CMD of KMHK~1762. Dots represent stars that remain after {\it Gaia} parallaxes and PMs cuts. Note that the cross-match with the {\it Gaia} dataset was able to clean the CMD only for stars brighter than $g \sim 21$~mag. Crosses indicate stars that should not be SC members based on their parallaxes and/or PM estimates. The empty circle indicates a star likely non-member of the cluster but with statistically compatible PMs (see text). Squares are stars for which we do not have any membership information. The red, blue and green dotted lines represent the isochrone of a SSP with parameters adopted by \citet[][]{Geisler-1997}, and those estimated in this work, {\emph Run 1} and {\emph Run 2}, respectively.
    Black solid lines are isochrones of 4 Gyr and 10 Gyr showing the boundaries of the age gap. In the same panel, completeness curves built through artificial star tests in the inner 0.25\arcmin~from the cluster center, are displayed as dashed lines.
    \emph{Right:} Cyan dots are field stars taken at 1\arcmin~from the cluster centre and lying in an area 25 times larger to compare the CMD of KMHK1762 with that of a local field. In the background, the Hess diagram of all stars within the YMCA tile 5\_38 is displayed. Blue points represent best-fit isochrone from {\emph Run 1}. In the same panel we show the typical photometric uncertainties.}
    \label{fig:kmkh_lit}
\end{figure}

Figure~\ref{fig:kmkh_lit} (left panel) shows the CMD of KMHK~1762 where the stars that have parallaxes or PMs incompatible with those of the cluster are highlighted with different symbols.
The CMD presents a clump of stars at $g\sim$22.0 mag which we identify as the SC MSTO, a sub-giant branch (SGB) as well as a few stars in the red giant branch (RGB) and in the RC, which can be identified at $(g-i, g) \simeq (0.8, 19.5)$ mag.
In the right panel of the same Figure, we displayed the CMD of a field used as a comparison, by taking all stars within an annulus of inner radius $r_{\rm in} = 1.0\arcmin$ (i.e. two times larger than the KMHK~1762 estimated radius) having an area 25 times larger than that of the SC.
Normalizing the number of field stars to the cluster area, only a handful of stars within the cluster area are expected to be non-cluster members, preventing any effort to remove contamination from field and foreground stars through the CMD, as discussed above.\par
The left panel of Fig.~\ref{fig:kmkh_lit} shows the CMD of KMHK~1762 superimposed with an isochrone from the PARSEC database \citep{Bressan-2012}\footnote{Note that these models adopt Z$_{\odot} = 0.0152$} with the age and metallicity reported by \citet[][]{Geisler-1997}, that is $t \simeq 2.7$~Gyrs and $[{\rm Fe/H}] = -0.91$~dex (in addition they used $E(B-V) = 0.15$~mag, and $m - M = 18.49$ mag).
The figure shows clearly that the isochrone with the labelled age and metallicity does not match the SC stars on the CMD. In particular, the isochrone's SGB is about one mag brighter than the stars piled up at $g \sim 22$~mag and in the colour interval $g-i \simeq [0.2,1]$ which we assume to be the actual SGB of the SC. Therefore, the proper SC age should be much older than previously estimated.\par
We thus carried out an objective isochrone matching procedure through the Automated Stellar Cluster Analysis package \citep[\tt ASteCA,][]{Perren-2015}. {\tt ASteCA} adopts synthetic generated single stellar populations coupled with a genetic algorithm to find the isochrone that best matches the observed CMD.
We performed two runs with {\tt ASteCA} adopting two different choices for the priors. In particular, in \emph{Run 1} we fixed the metallicity to the value $[{\rm Fe/H}] = -0.91$~dex estimated through spectroscopic measurements by \citet{Olszewski-1991}, while in \emph{Run 2} we let it free to vary in the $-3 \leq [{\rm Fe/H}] \leq 0$~dex interval. In both runs we restricted the distance modulus between 18 and  19 mag, which encompasses the average LMC distance of $m - M = 18.49$~mag \citep{deGrijs&Wicker&Bono2014}.
In Table~\ref{tab:kmhk1762_properties} we listed the output of {\tt ASteCA} in the two configurations, while the best isochrones from each of the two runs are overlaid to the KMHK~1762 CMD in Fig.~\ref{fig:kmkh_lit} (left panel).
The isochrone calculated with \emph{Run 1} (i.e. with fixed metallicity) better approximates the bright end of the RGB compared with that from \emph{Run 2}, which shows a more bended RGB at $g \leq 19.5$~mag, as a consequence of the larger estimated metallicity (i.e. [Fe/H]=$-0.65^{+0.27}_{-0.41}$~dex). Even the RC, identified at $(g-i, g) \simeq (0.8, 19.5)$ is better matched with \emph{Run 1} set of parameters, whereas the isochrone obtained in \emph{Run 2} provides a slight fainter ($g \sim 20$~mag) RC magnitude with respect to the observed CMD.
We therefore judge that the solution provided by \emph{Run 1} is the best one, even if, from the purely statistical point of view the two runs provide parameters in agreement with each other within the uncertainties.
The left panel of Figure~\ref{fig:kmkh_lit} also shows completeness curves in the inner 0.25\arcmin ~from the cluster center, derived as described in Sect.~\ref{sec:data}. The putative SGB is entirely at the 80\% completeness level, giving us confidence that the actual cluster age is well within the uncertainties provided by {\tt ASteCA}\footnote{Beyond 0.25\arcmin~from the cluster center, the corresponding completeness levels are about 0.5-0.75 mag deeper.}, and thus is confidently within the age gap, as also shown by the isochrones of 4 Gyr and 10 Gyr (boundaries of the age gap) in the left panel of Figure~\ref{fig:kmkh_lit}.\par
\par
Our best age estimate is therefore considerably larger compared with the \citet{Geisler-1997} estimate. This discrepancy can be due to the different kind of data and analysis between their and our work. 
To estimate the age of KMHK~1762, \citet{Geisler-1997} did not use the isochrone fitting method (note that the PARSEC isochrones were not available at that time), but adopted a calibration of the magnitude difference between the RC and MSTO vs age in the Washington photometric system. They observed in the Washington photometric system with a shallower faint limit ($T1\sim21-22$ mag) and a consequent significant noise at the level of MSTO, which makes it difficult to identify precisely this feature.
In the right panel of Figure~\ref{fig:kmkh_lit} we also displayed the best isochrone retrieved in \emph{Run 1} to show that the bulk of the LMC stellar population within the YMCA tile 5\_38 is older than KMHK~1762, and therefore the SGB we observe at g $\sim$ 22 mag is unlikely a contamination effect by LMC field stars.


\begin{table}
    \caption{Properties of KMHK 1762.}
    \label{tab:kmhk1762_properties}
    \centering
    \large
    \begin{tabular}{l|c|c}
    \hline \hline 
    Property & Run 1 & Run 2 \\
    \hline 
    $\log (t)$ & $9.74 \pm 0.15$~dex & $9.69 \pm 0.14$~dex\\
    &&\\[-0.8em]
    $\mu_0$ & $18.62^{+0.24}_{-0.29}$~mag & $18.68^{+0.19}_{-0.27}$~mag\\
    &&\\[-0.8em]
    $[$Fe/H$]$ & $-0.91$~dex (fixed) & $-0.65^{+0.27}_{-0.41}$~dex\\
    &&\\[-0.8em]
    E(B-V) & $0.09^{+0.06}_{-0.05}$~mag & $0.08^{+0.07}_{-0.05}$~mag\\
    \hline
    \end{tabular}
     \tablefoot{The errors were provided by the {\tt ASteCA} package as 16-th and 84-th around the median value.}

\end{table}

\section{Discussion}

\begin{figure}
    \hspace{-0.2cm}
    \includegraphics[width=0.5\textwidth]{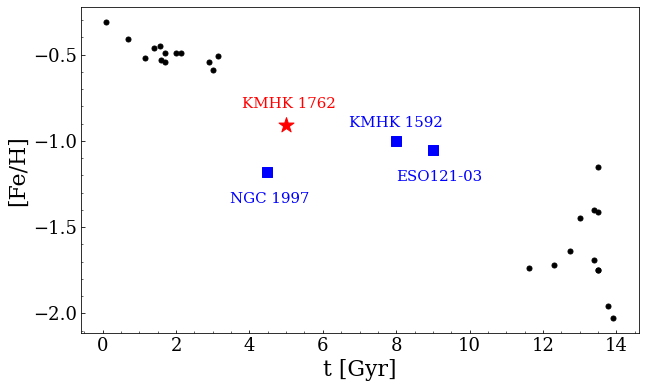}
    \caption{Metallicity as a function of the age for the LMC analyzed by \citet{Song-2021} and \citet{Mucciarelli-2021} is plotted as black points.
    The red star marks the position in the plot of KMHK~1762, while blue squares indicate the SC ESO121-03 and KMHK 1592. For KMHK~1762, we adopted the metal-content estimated by \cite{Olszewski-1991} (i.e. Fe/H = -0.91 dex).}
    \label{fig:kmhk_amr}
\end{figure}

The age of KMHK~1762 estimated as discussed in the previous section is t=$5.5^{+2.3}_{-1.6}$ Gyr. This means that it is the third confirmed age gap SC, in addition to ESO121-03 and KMHK~1592, as mentioned in the introduction.  Hereafter we also include NGC~1997 in the discussion as the end of the age gap period is not strongly constrained and this SC may also fall within it. 
The actual presence of an age gap among the LMC SC system is still an open question, considering that both old and recent studies devoted to the reconstruction of the LMC SFH in the stellar field do not show an absence of star formation in the $\sim$4-10 Gyr interval \citep[see e.g.][and references therein]{Tosi2004,Piatti&Geisler2013,Rubele-2012,Mazzi-2021}.
More quantitatevely, adopting the recent work by \citet{Mazzi-2021} we can calculate that the LMC formed at least $0.5 \times 10^9 M_{\odot}$ in the age gap period (see their figure 17).
Therefore, the hypothesis of a quiescent period of SC formation in the LMC during the age gap interval contradicts one of the paradigms of star formation, which foresees that a great fraction of stars is formed within SCs \citep{Lada&Lada2003}.
Hence, we should conclude that we do not observe SCs in the age gap because they could have been destroyed by the tidal forces of the LMC, but this hypothesis is at odds with the steep increase of the SC age distribution at $\sim$ 3 Gyr \citep[see e.g.][G20, and references therein]{Pieres-2016}. Even assuming that this sharp increase in SC formation at about 3 Gyr ago is due to a close encounter with the SMC, it is however difficult to imagine an ad hoc disruption mechanism which only act in the age interval 4--10 Gyr and becomes suddenly inefficient at an age of 3 Gyr \citep{DaCosta2002}.\par
The presence of the age gap also prevents the derivation of the age-metallicity relation (AMR) by means of the LMC SC system, and furthermore a similar gap is present also in the metallicity \citep{Rich-2001}. 
The AMR for a subsample of LMC SCs is shown in Fig.~\ref{fig:kmhk_amr}. In particular, we adopted the samples recently analysed through spectroscopic observations by \citet{Song-2021} and \citet{Mucciarelli-2021}.
While the old LMC GCs are on average all metal-poor (the majority has $[{\rm Fe/H}] \leq -1.3$~dex), the younger SCs have on average $[{\rm Fe/H}] \simeq -0.7$~dex, thus they are separated by a considerable gap in the metal content.
In between the two sub-populations stands out the presence of ESO121-03, KMHK~1592, and now KMHK~1762, thanks to its newly estimated age. Note the position of NGC~1997 which appears too young for the estimated metallicity, which is however not derived from spectroscopy, but by isochrone fitting to a CMD.\par
The case of KMHK~1762 appears emblematic: SC ages estimated on the basis of too shallow photometry or not correctly de-contaminated CMDs can result significantly younger than their true age value. It is worth noticing that also in the case of NGC~1997, the analysis of deeper data led to an age estimate 2 Gyr older than past investigations, shifting this SC slightly into the age gap \citep{Pieres-2016}.
These results do not clarify the physical origin of the age gap, supporting instead its explanation as an observational bias.
Indeed, G20 showed that, while the spatial distribution of young SCs traces the main features of the LMC, such as the central bar or the spiral arms, that traced by SCs older than 1 Gyr is rather clumpy (see their Fig. 18), generally resembling the regions explored with modern deep photometric observations.
In fact, most of the works devoted to the search of undiscovered SCs focused in the LMC central regions leaving the outskirts (i.e. $d > 4$~kpc) quite unexplored and were conducted on the basis of photometrically shallow surveys, allowing the researchers to detect only SCs younger than $\sim$~1–1.5 Gyrs \citep{Pietrzynski&Udalski2000,Glatt-2010,Nayak-2016}.\par
\begin{figure}
    \includegraphics[width=\hsize]{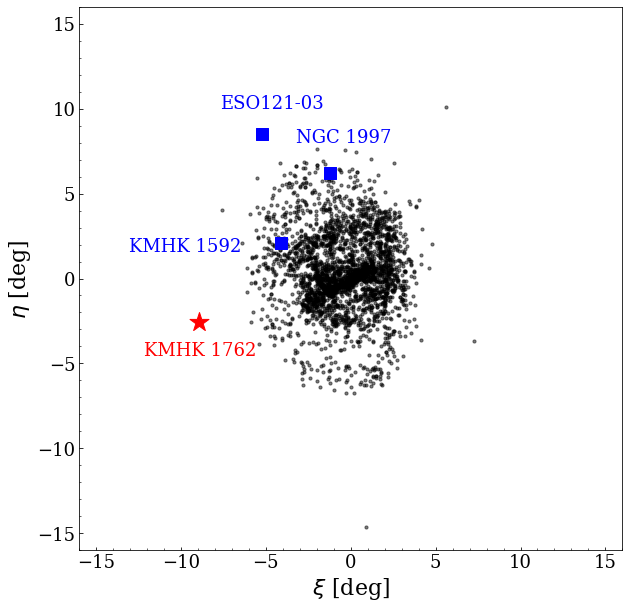}
    \caption{Relative positions of SCs collected in the \citet{Bica-2008}'s catalogue with respect to the LMC centre. The red star indicates the position of KMHK~1762, whereas the blue squares mark the position of ESO121-03, KMHK~1592, and NGC~1997, the other age gap SCs known hitherto.}
    \label{fig:kmhk1762_positions}
\end{figure}

Fig.~\ref{fig:kmhk1762_positions} displays the relative positions of all SCs collected in the \citet{Bica-2008}'s catalogue with respect to the LMC centre. All age gap SCs discussed in this work reside in the outer regions of the LMC, indicating that we can more easily detect them in the periphery, either because of fainter tidal stresses that let them survive longer (physical effect) or as a consequence of a less crowded environment (observational effect).
These arguments suggest that the issue of the SC age gap in the LMC deserves to be revisited (a) after a more complete census of the SC population has been obtained (Gatto et al., in prep.), and (b) with new age estimates based on deep photometry for known (supposedly) intermediate age clusters.
Both actions will certainly be possible in the near future, once the Rubin-LSST survey will enter into operation and provide the first results. 

\section{Summary}

In this work we presented a study of the SC KMHK~1762 based on the deep photometry provided by the YMCA survey which allowed us to construct a CMD in the $g$ and $i$ bands reaching at least 1.5-2 mag below the MSTO, that is significantly deeper than the previous works in the literature. 
We took advantage of the PMs and parallaxes provided by the {\it Gaia} EDR3 to mitigate the contamination of the KMHK~1762 CMD due to LMC and MW foreground/background stars.
The bright limiting magnitude of the {\it Gaia} mission (i.e. $g \simeq 21$ mag) allowed us to clean the CMD from non-cluster members only in the post-main sequence evolutionary phases, that are however crucial to constrain the metal-content of the SC based on the inclination of the RGB.\par
The SGB clearly visible at $g \sim 22$ mag indicates that KMHK 1762 is older than previously estimated based on shallower photometry.
Indeed, the automatic isochrone fitting procedure performed with the {\tt ASteCA} package yielded an age of $t = 5.5^{+2.3}_{-1.6}$ Gyr, and analysis of completeness levels suggest is a lower limit for the age, making it the third or the fourth confirmed age gap SC ever discovered, after ESO121-03 and KMHK 1592, and possibly also NGC~1997.
We speculate that other intermediate-age SCs analysed with shallow photometry could actually be older and thus potentially formed during the age gap period.
In addition, as recently shown by G20, several age gap SCs could be hidden in the outskirts of the LMC, and due to their intrinsic faintness, can only be revealed on the basis of deep photometry. 
On these grounds, the increased number of confirmed or suspected SCs formed in the age gap period suggests that the age gap may be an observational bias, possibly combined with a high destruction rate in the more LMC central regions, rather than a true quiescent period of SC formation in the LMC.

\begin{acknowledgements}
We warmly thank the anonymous referee for the helpful comments and suggestions that greatly improved the manuscript and helped to strengthen the results of  this work.
M.G. and V.R. acknowledge support from the INAF fund "Funzionamento VST" (1.05.03.02.04).
This work has been partially supported by INAF through the “Main Stream SSH program" (1.05.01.86.28).
This work is based on observations collected at the ESO within the VST Guaranteed Time Observations, Programme ID: 0105.D-0198.

\end{acknowledgements}

%
\bibliographystyle{aa} 
\bibliography{unatesi} 
%

\end{document}